\documentclass[12pt]{article}
\input epsf
\usepackage{amsmath}
\usepackage{amssymb}
\usepackage[dvips]{graphicx,psfrag}
\usepackage{lscape}
\usepackage{subfloat}
\newcommand{\be}{\begin{equation}}
\newcommand{\ee}{\end{equation}}
\newcommand{\bea}{\begin{eqnarray}}
\newcommand{\eea}{\end{eqnarray}}
\newcommand{\lb}{\label}

\begin{document}
\begin{titlepage}
\title{Scalar-Tensor Models of Normal and Phantom Dark Energy}
\author{Radouane Gannouji\thanks{email:gannouji@lpta.univ-montp2.fr}~,
David Polarski\thanks{email:polarski@lpta.univ-montp2.fr}~,
Andr\'e Ranquet\thanks{email:ranquet@lpta.univ-montp2.fr}\\
\hfill\\
Lab. de Physique Th\'eorique et Astroparticules, UMR 5207\\
Universit\'e Montpellier II, 34095 Montpellier Cedex 05, France\\
\hfill\\
Alexei A. Starobinsky \thanks{email:alstar@landau.ac.ru}\\
\hfill\\
Landau Institute for Theoretical Physics, Moscow, 119334, Russia}
\pagestyle{plain}
\date{\today}

\maketitle

\begin{abstract}
We consider the viability of dark energy (DE) models in the
framework of the scalar-tensor theory of gravity, including the
possibility to have a phantom DE at small redshifts $z$ as
admitted by supernova luminosity-distance data. For small $z$, the
generic solution for these models is constructed in the form of a
power series in $z$ without any approximation. Necessary
constraints for DE to be phantom today and to cross the phantom
divide line $p=-\rho$ at small $z$ are presented. Considering the
Solar System constraints, we find for the post-Newtonian
parameters that $\gamma_{PN}<1$ and $\gamma_{PN,0}\approx 1$ for
the model to be viable, and $\beta_{PN,0}>1$ (but very close to
$1$) if the model has a significantly phantom DE today. However,
prospects to establish the phantom behaviour of DE are much better
with cosmological data than with Solar System experiments. Earlier
obtained results for a $\Lambda$-dominated universe with the
vanishing scalar field potential are extended to a more general DE
equation of state confirming that the cosmological evolution of
these models rule them out. Models of currently fantom DE which
are viable for small $z$ can be easily constructed with a constant
potential; however, they generically become singular at some
higher $z$. With a growing potential, viable models exist up to an
arbitrary high redshift.
\end{abstract}

\end{titlepage}

\section{Introduction}
A major turning point in cosmology has been reached with the  {\it
observational} discovery that our Universe is accelerating now
(and has been accelerating for several billion years in the past)
\cite{R98,P99}. If interpreted in terms of the Einstein equations
for the evolution of a Friedmann-Lema\^itre-Robertson-Walker
(FLRW) cosmological models with the (practically) zero spatial
curvature (the latter follows from other arguments), this means
that approximately two thirds of the total present energy density
of matter in our Universe is due to some gravitationally
unclustered component called Dark Energy (DE).

Observations of high redshift supernova, fluctuations of the
cosmic microwave background (CMB) temperature and other effects
tell us that the effective energy density $\rho_{DE}$ of DE is
very close to minus its effective pressure $p_{DE}$ (see Eqs.
(\ref{E1a},\ref{E2a}) below for the exact, though conventional,
definition of these quantities) and that they are both very weakly
changing (if at all) with time and with the expansion of the
Universe. The physical nature of DE is unknown at present, and
three main logical possibilities exist (see
\cite{SS00,PR03,Pad03,Sah05,CST06} for reviews).\\
1. DE is a cosmological constant, as originally suggested by
Einstein, with $\rho_{DE}= -p_{DE}=\Lambda/8\pi G=const$
exactly.\footnote{$\hbar=c=1$ is used throughout the paper.}\\
2. Physical DE: DE is the energy density of some new, very weakly
interacting physical field (e.g., a quintessence -- a scalar field
$\phi$ with some potential $V(\phi)$ minimally coupled to gravity). \\
3. Geometric DE: the Einstein general relativity (GR) equations
are {\em not} the correct ones for gravity, but we write them in
the Einsteinian form by convention, putting all arising additional
terms into the r.h.s. of the equations and calling them the
effective energy-momentum tensor of DE.

Of course, this classification is not absolute. In some cases, the 
difference between physical and geometric DE can, in turn, become
conventional. E.g., for a non-minimally coupled scalar field 
(which constitutes a specific case of scalar-tensor gravity considered
in this paper), equations of the model have the same form irrespective
of the origin (physical or geometric) of this field. Another way of
classifying DE models which is more invariant and, therefore, more 
important from the point of view of the observational determination of 
the type of DE is to divide these models into dynamical, if the DE 
description requires a new field degree of freedom ($\equiv$ a new 
particle from the quantum point of view), and non-dynamical in the 
opposite case. Then all physical DE models and many of the geometric 
ones belong to the first class, while the cosmological constant itself 
and some geometric models fall into the second category, i.e., the $F(R)$ 
model with the Palatini variation of its action where $R$ is the scalar 
curvature.

At present, all existing observational data are in agreement with
the simplest, first possibility (inside $\sim 2\sigma$ error bars
in the worst case). The case of a cosmological constant is
internally self-consistent and non-contradictive. The extreme
smallness of the cosmological constant expressed in the either
Planck, or even atomic units means only that its origin is {\em
not} related to strong, electromagnetic and weak interactions (in
particular, to the problem of the energy density of their vacuum
fluctuations) at all. However, in this case we remain with one
dimensionless constant only and can not say anything more (at
least, at present). That is why the two other possibilities
admitting a (slightly) variable dark energy have been also
actively studied and compared with observational data recently.
Moreover, properties of the present DE are remarkably {\em
qualitatively} similar to those of an "early DE" that supported an
inflationary stage in the early Universe. But in the latter case,
we are sure that this early DE was unstable. So, it is natural to
conjecture by analogy that the present DE is not stable, too.

On purely phenomenological grounds, one can consider DE models
with a constant equation of state parameter $w_{DE}\equiv p_{DE}/
\rho_{DE}$ different from $-1$. However, the latest observational
data have already severely restricted this simplest alternative
possibility to $|1+ w_{DE}|\stackrel {<}{\sim} 0.1$ ($1\sigma$
error bars) \cite{Sel05,As06,SPD06}. Therefore, a viable
alternative to the cosmological constant has to be looked for
among more complicated models with $w_{DE}\not= const$. Such
models include quintessence ones with different potentials
$V(\phi)$ (see \cite{SS00,PR03,Pad03,Sah05} for numerous
references), models with several minimally coupled scalar fields
(see, e.g., \cite{BP04}), those with direct non-gravitational
coupling between DE and dark matter (DM) (\cite{A00} and following
papers), unified models of DE and DM (the Chaplygin gas model
\cite{Kam01} and others), etc. Then, however, it becomes very
important to investigate if there exist models of DE for which a
variable $w_{DE}$ may cross the "phantom divide" line $w_{DE}=-1$
(we call DE "fantom" if it has $w_{DE}(z) <-1$ for a given $z$ and
"normal" in the opposite case). Note that the weak energy
condition (WEC) is violated for phantom DE. This is not possible
at all for quintessence models with the standard kinetic term and
hardly possible, i.e., it requires non-generic initial conditions,
for scalar field models having a non-standard kinetic term
("$k$-essence") \cite{Vik05} (see also \cite{ACK05}).

Indeed, analysis of the recent SNe data with redshifts up to
$z=1.7$ (the "Gold dataset" \cite{Rie04}) using fits containing at
least 2 free parameters, e.g., the linear fit (\ref{wCPL}) for
$w_{DE}$ in terms of the scale factor $a$ \cite{CP01,L03} or the
quadratic polynomial fit (\ref{wp}) for $\rho_{DE}$ in terms of
the redshift $z$ \cite{SSSA03}, or with model independent methods,
results in best fits to these data having a variable $w_{DE}$ that
steadily increases for redshifts $0<z<1$ and crosses the "phantom
divide" somewhere between $0$ and $0.5$ \cite{ASS04,HC05,Gong05,
DR04,HM04,RAW05,UIS05,LNP05,ER05,XZF06} (see, however, \cite{WT,
Sel05,JBP06,DKC06} for a more conservative view -- in the sense
of returning back to a cosmological constant). This statement does
not mean that an exact cosmological constant is excluded -- it is
still inside $1$ or $2\sigma$ error bars, as was stated above.
Moreover, one should be cautious with this result: it may be a
consequence of trying to obtain a too fine-grained graph of
$w_{DE}(z)$ from the luminosity distance $d_L(z)$ as has been
already emphasized in \cite{MBMS02}. In this respect, results for
$w_{DE}(z)$ averaged over a range of redshifts $>0.3$ (dubbed the
"$w$-probe" in \cite{SASS05}) are more statistically reliable,
e.g., the result that $\bar w_{DE}\approx -1.05_{-0.07}^{+0.09}$
obtained for the same Gold SNe dataset in \cite{ASS04} for
$\Omega_{m,0}=0.3$ and the averaging redshift interval
$(0,0.414)$. However, a remarkable possibility of crossing the
phantom divide at recent redshifts, when DE has become the
dominant component of the Universe by its effective energy
density, still remains viable. The latest Supernovae data with
$z\sim 0.5$ \cite{CSF05} also admit $w_{DE}(z)<-1$ for $z<0.5$,
but not for larger redshifts, as a possible interpretation.
Finally, the recent data on acoustic baryon oscillations in the
present matter power spectrum (the Sakharov oscillations)
\cite{Eis05,Col05}, while strongly restricting one direction in
the plane of parameters for the two-parametric fits
(\ref{wCPL},\ref{wp}), leave the orthogonal direction practically
unconstrained, therefore, permitting the recent phantom divide
crossing (see Sec.~6).

So, if this striking behaviour of DE will be confirmed by future
data, what is the best way to describe it? One possibility, a {\em
ghost} phantom DE, i.e. a scalar field with the negative sign of
its kinetic term, was first proposed in \cite{Cald02} and
triggered a very large wave of publications (see, e.g., the recent
papers \cite{NOT05,STTT05} for a list of references on this
topic). However, it has been long known that theories of this type
are plagued by quantum instabilities, the most dangerous of those
being the process of creation of two particle+antiparticle pairs:
one of the ghost field and another of any usual (non-ghost) field
(see \cite{CJM04} for a recent investigation). Moreover, a ghost
model of DE is unsatisfactory even at the classical level: it does
not explain the observed large-scale isotropy and homogeneity of
the Universe! Just the opposite, e.g., for the given Hubble
constant $H_0$ averaged over angular directions in the sky, we
would expect a universe to be very strongly anisotropic with the
anisotropy energy density (i.e., the positive energy density of
long-wavelength gravitational waves) being compensated by a large
negative energy density of the ghost DE, or of a ghost component
of DE in more complicated multi-component models of this type
(e.g., in the two-field realization of the ``quintom'' model
introduced in \cite{FWZ05}). It is just a classical analogue of
the quantum instability mentioned above, with the "usual" field
being the gravitational one and with the dynamical quantum
instability transformed into the problem of classical initial
conditions. For this reason, we are sceptical regarding this
approach as a whole.

 Fortunately, there is no necessity to introduce a ghost field to
explain possible phantom behaviour of DE including its phantom
divide crossing. As was first emphasized in \cite{BEPS00},
scalar-tensor theories of gravity allow for this phenomenon.
Scalar-tensor models of DE belong to the third class (geometric
DE) according to the classification given above. This class of
models is very rich and interesting. It contains the Einstein
gravity plus a non-minimally coupled scalar field, as well as the
higher-derivative $f(R)$ gravity, where $R$ is the scalar
curvature, as particular cases, besides allowing for DE phantom
behaviour and transition to a normal DE. In this paper, we will
concentrate on scalar-tensor DE models with the latter properties
because of tentative observational evidence described
above.\footnote{The third possibility to get a phantom DE
including the phantom divide crossing which is based on braneworld
models \cite{SSh03} (see also the recent paper \cite{CLMQ06}) will
not be discussed here.} Note that a possible phantom behaviour of
DE in scalar-tensor gravity has a conventional character. The
reason is that the effective gravitational constants $G_N$ and
$G_{\rm eff}$ (see Sec.~2 below) are generically time-dependent
while the definitions (\ref{E1a},\ref{E2a}) of the DE effective
energy density and pressure assume some constant $G$ when writing
the left-hand side of equations and splitting their right-hand
side into energy-momentum tensors of non-relativistic matter
(mainly non-baryonic cold dark matter) and dark energy. As a
result, a part of the Einstein tensor $G_{\mu\nu}$ and the
energy-momentum tensor of dark matter multiplied by a change in
$G_N$ is conventionally attributed to DE. In other words, phantom
behaviour of scalar-tensor DE is always 'curvature-induced', in
contrast to the ghost DE models or other models like those
considered in \cite{AKV05,Rub06} where it may occur in the flat
space-time already.

The possibility to get both a phantom DE and the phantom divide
crossing in scalar-tensor gravity is related to the fact that this
theory has two arbitrary and independent functions $F(\Phi)$ and
$U(\Phi)$ (see the Lagrangian (\ref{L}) below). Throughout this
work we assume spatial flatness though this prior while well
motivated theoretically can be challenged \cite{PR05}. As has been
shown in \cite{BEPS00}, two different types of observations, e.g.,
determination of the luminosity distance and the inhomogeneity
growth factor in the non-relativistic component as functions of
redshift, are necessary and sufficient for the total
reconstruction of the microscopic Lagrangian (\ref{L}) of
scalar-tensor gravity. However, as shown in \cite{EP01}, a partial
reconstruction using the luminosity distance data only could yield
interesting information, too. In that case some assumption about
one of the functions $F(\Phi)$ and $U(\Phi)$ has to be made, so
that only one unknown function in the microscopic Lagrangian
(\ref{L}) remains to be found. Clearly, many different partial
reconstruction strategies are possible and we explore some of them
here with the aim to investigate which DE models in scalar-tensor
gravity are viable.

It has been found in \cite{EP01} that $\Lambda$-dominated
universes with a vanishing potential $U$ are ruled out as they
lead to singular universes already at very low redshifts $z\sim
0.7$ (see also the recent paper \cite{Per05}). 
Of course, completely regular but non-accelerating solutions 
do exist in this case.
Among models with a
non-zero potential $U$, one concrete example of a model with the
phantom divide crossing was constructed in \cite{Per05} where the
functions $F(\Phi)$ and $U(\Phi)$ were given in the parametric
form as functions of $z$ up to $z\approx 3$. More examples for a
non-minimally coupled scalar field ($F(\Phi)=F_0-\xi\Phi^2$) with
a non-zero potential $U$ were investigated in \cite{LS05}, also for
$z<2$. In the present paper, we make a next step and extend these
results in two directions: first, by constructing a generic solution
for scalar-tensor DE models for $z<1$ in the form of a power series
in $z$; second, by investigating and numerically integrating some of
these solutions up to large $z\gg 1$. The latter task appears to be
necessary since, though DE is subdominant for $z\gg 1$ as compared
to non-relativistic non-baryonic dark matter and baryons, its model
itself may become intrinsically contradictory for large $z$, namely,
$F$ or $\dot\Phi^2$ may become negative for an unfortunate choice of
$U(\Phi)$. It is also crucial to check that any DE model has the 
correct power-law behaviour for large $z$ \cite{APT06}.

Finally, for the scalar-tensor theory of gravity, it is well known
that the present value of ${dF\over d\Phi}(z=0)$ is severely
restricted from Solar System tests of post-Newtonian gravity (i.e.
by the measured values of the post-Newtonian parameters). For this
reason, there have been stated that the present phantom behaviour
in scalar-tensor models of DE requires large amount of fine tuning
and, thus, is unnatural \cite{T02,SFT05}. Therefore, it is
important to investigate this problem in more detail to quantify
what amount of fine tuning (and of what kind) is necessary for a
significantly phantom behaviour of DE, and to determine the
relation between this behaviour and results of Solar System tests
of gravity.

The paper is organized as follows. In Sec.~2 we define all
quantities related to our scalar-tensor DE model and present the
background evolution equations. 

In Sec.~3 we derive the general integral solution for the quantity 
$F(z)$. In Sec. 4 we consider solutions in which DE scales as some power 
of the FLRW scale factor $a(t)$ and show their existence for DE of the
phantom type (the latter requires a non-zero $U(\Phi)$).

In Sec.~5 we consider the small $z$ behaviour of our model and find
conditions for the violation of the WEC by DE today and for the
phantom boundary crossing at small $z$. In Sec.~6 the general
reconstruction of a microscopic model is considered and the
observational constraint from the acoustic oscillations in the
matter power spectrum is discussed. In Sec.~7 we consider the
reconstruction for a constant potential more specifically and show
that the model becomes singular at some redshift that cannot be
arbitrarily high once we have chosen a specific equation of state
$w_{DE}$. In Sec.~8 non-constant potentials are considered and a
model which is asymptotically stable for large $z$ is presented.
Sec.~9 contains conclusions and discussion.

\section{Background evolution}
In this section we review the background evolution equations in a
spatially flat FLWR universe. We consider a model where gravity is
described by a scalar-tensor theory and we start with the
following microscopic Lagrangian density in the Jordan frame
\begin{equation}
L={1\over 2} \Bigl (F(\Phi)~R -
Z(\Phi)~g^{\mu\nu}\partial_{\mu}\Phi\partial_{\nu}
\Phi \Bigr) - U(\Phi) + L_m(g_{\mu\nu})~.
\label{L}
\end{equation}
Since $L_m$ is not coupled to $\Phi$, the Jordan frame is the
physical one. In particular, fermion masses are constant and
atomic clocks measure the proper time $t$ in it. The quantity
$Z(\Phi)$ can be set to either $1$ or $-1$ by a redefinition of
the field $\Phi$, apart from the exceptional case $Z(\Phi)\equiv
0$ when the scalar-tensor theory (\ref{L}) reduces to the
higher-derivative gravity theory $R+f(R)$. In the following, we
will write all equations and quantities for the case $Z=1$. For
our purposes, $L_m$ describes non-relativistic dust-like matter
(baryons and cold dark matter) as we are interested in low
redshift ($z\ll z_{eq}$) behaviour only. Here, $z_{eq}$ denotes
the equality redshift when the energy densities of
non-relativistic matter and radiation are equal. In such a model,
the effective Newton gravitational constant for homogeneous
cosmological models is given by
\be
G_N= (8\pi F)^{-1}.
\ee
As could be expected, $G_N$ does not have the same physical meaning
as in General Relativity, the effective gravitational constant
$G_{\rm eff}$ for the attraction between two test masses is given
by
\be
G_{\rm eff} = G_N ~{F+2(dF/d\Phi)^2\over
F+\frac{3}{2}(dF/d\Phi)^2}~.\lb{Geff}
\ee
on all scales for which
the field $\Phi$ is effectively massless \cite{BEPS00} and $F>0$.
The condition $G_{\rm eff}>0$ is one of the stability conditions
of the scalar-tensor theory (\ref{L}), it means that the graviton
is not a ghost. In fact, even at the purely classical level, it
has been shown in \cite{St81} that a generic solution of (\ref{L})
may not smoothly cross the boundary $G_{\rm eff}=0$. Instead, a
curvature singularity forms at this boundary which generic
structure has been also constructed in \cite{St81}. This condition
combined with another stability condition (see Eq. (\ref{posEn1})
below) results in $F>0$, so $G_N>0$, too.

We will write all equations in the Jordan frame using (\ref{L}).
Specializing to a spatially flat FLRW universe
\be
ds^2 = - dt^2 + a^2(t)~d{\bf x}^2~,
\ee
the background equations are as follows:
\bea
3FH^2 &=& \rho_m + {\dot \Phi^2\over 2} + U - 3H {\dot F}~,\lb{E1}\\
-2 F {\dot H} &=& \rho_m + \dot \Phi^2 + {\ddot F} - H {\dot F}~.\lb{E2}
\eea
The evolution equation of the scalar field $\Phi$ can be obtained
from the two equations (\ref{E1},\ref{E2}). Eliminating the quantity
$\dot \Phi^2$ by combining these equations, we obtain a master
equation for the quantity $F$ which takes the following form when all
quantities are expressed as functions of redshift $z$:
\begin{eqnarray}
F'' &+& \left[(\ln h)' - \frac{4}{1+z}\right]~F' +
\left[\frac{6}{(1+z)^2} - \frac{2(\ln h)'}{1+z}\right]~F
\nonumber\\
&=& \frac{6 u}{(1+z)^2~h^2} F_0 ~\Omega_{U,0} + 3~(1+z)~h^{-2} F_0~
\Omega_{m,0}\ ,\lb{MF}
\end{eqnarray}
where a prime denotes the derivative with respect to $z$ and we have
introduced the quantities $h\equiv\frac{H}{H_0}$, $\Omega_{U,0}\equiv
\frac{U_0}{3F_0~H_0^2}$ and $u\equiv\frac{U}{U_0}$. The index $0$
denotes the present moment here and below. The (dimensionless)
relative energy density $\Omega_{m,0}$ is defined through
$\Omega_{m,0}\equiv \frac{\rho_{m,0}}{3F_0H_0^2}$. Once the master
equation (\ref{MF}) is solved for $F$, we get the algebraic equation
for $\Phi'(z)$:
\be
\frac{\Phi'^2}{6} = -{F'\over 1+z} + {F\over (1+z)^2}
-\frac{F_0~u}{(1+z)^2~h^2} ~\Omega_{U,0}
-\frac{F_0~(1+z)}{h^2} ~\Omega_{m,0}~.\lb{Phi}
\ee

The second stability condition of the scalar-tensor gravity (\ref{L})
is
\be
\omega_{BD} = \frac{F \Phi'^2}{F'^2} > -\frac{3}{2}~,\lb{posEn}
\ee
where $\Phi'^2$ is found from (\ref{Phi}).
Inequality (\ref{posEn}) just expresses the positivity of the energy
of the (helicity zero) scalar partner of the graviton, i.e. the
positivity of the kinetic energy of the scalar field
in the Einstein frame (see e.g. \cite{EP01} for more details)
\be
\phi'^2 \equiv {3\over 4}\left({F'\over F}\right)^2 + {\Phi'^2
\over 2F} > 0~
\lb{posEn1}
\ee
with $\Phi'^2$ taken from (\ref{Phi}).
With the $Z=1$ parametrization, we cannot reconstruct the function
$F(\Phi)$ when $-\frac{3}{2}<\omega_{BD}<0$ since $\Phi'^2$ becomes
negative in this case. Indeed, these allowed negative values of
$\omega_{BD}$ correspond to the parametrization choice $Z=-1$ in
Eq.(\ref{L}). The $Z=1$ parametrization allows us to consider
consistently only cases for which $\Phi'^2\ge 0$, or equivalently
$\omega_{BD}\ge 0$. However, the condition (\ref{posEn}) with
$\Phi'^2$ given by (\ref{Phi}) is the true condition that the theory
is well behaved and it remains valid even for $\Phi'^2<0$. This is
best understood in the Brans-Dicke parametrization $F=\Phi,~Z=
\frac{\omega_{BD}}{\Phi}~$, where one can reconstruct the two
functions $F>0$ and $\omega_{BD}>-\frac{3}{2}$ from (\ref{MF}),
(\ref{Phi}), (\ref{posEn}). For $\Phi'^2\ge 0$ and $F\ge 0$, the
inequality (\ref{posEn}) is satisfied automatically. The inclusion
of the range $-\frac{3}{2}<\omega_{BD}<0$ is not of purely academic
interest and can be important when one considers the reconstruction
of DE models far in the past. We illustrate this with a specific
example in Fig. 1.

\begin{figure}[ht]
\begin{center}
\psfrag{F}[tl][br][1.5]{} \psfrag{z}[][][1.5][0]{$z$}
\includegraphics[angle=-90,width=1\textwidth]{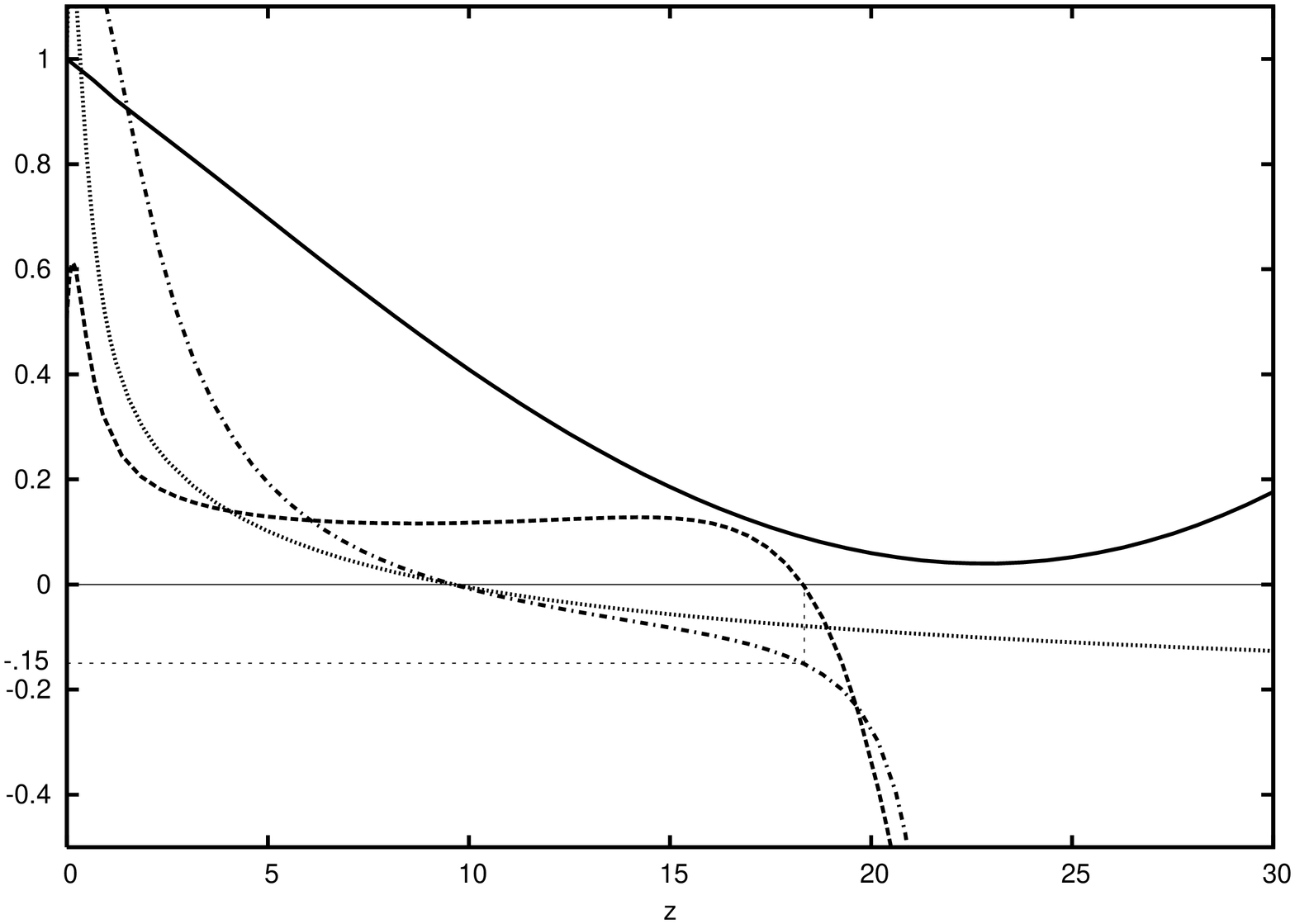}
\end{center}
\caption[]{Several quantities are displayed for the model with parametrization (\ref{wCPL})
with $\alpha\equiv 1+w_0=-0.2$ and $\beta \equiv w_1=0.4$, while $F_1=0$ and
$\frac{\Omega_{U,0}}{\Omega_{DE,0}}=\frac{\Omega_{U,0}}{0.7}=0.97585$.
The curves shown represent the following (rescaled)
quantities in function of redshift $z$: $\frac{F}{F_0}$ (solid), $10\times \Phi'^2$ (dotted),
$10\times \phi'^2$ (dashed) and $0.1\times \omega_{BD}$ (dot-dashed). It is seen that
$\omega_{BD}$ and $\Phi'^2$ become negative at $z\approx 10$. The model remains valid beyond
$z\approx 10$ untill  $z\approx 18$ as long as $\omega_{BD}>-\frac{3}{2}$ or equivalently
$\phi'^2>0$. So, we see that the model remains valid for a large interval where $\Phi'^2<0$.
Of course, it is impossible to reconstruct $\Phi$ in this interval using the $Z=1$
parametrization. Note that $\phi'^2$ becomes negative {\it before} $z_m\approx 22$ where
$F'(z_m)=0$, in accordance with (\ref{ineq}).}
\end{figure}

Solar System experiments constrain the post-Newtonian parameters
$\gamma_{PN}$ and $\beta_{PN}$ {\it today} (for these quantities,
we drop here the subscript $0$)
\bea
\gamma_{PN} &=& 1 - {(dF/d\Phi)^2\over F + 2(dF/d\Phi)^2}~,
\lb{gammaPPN}\\
\beta_{PN} &=& 1 + {1\over 4} {F~(dF/d\Phi)\over 2 F +
3(dF/d\Phi)^2}{d\gamma\over d\Phi}~,\lb{betaPPN}
\eea
as well as the quantity $\frac{{\dot G}_{{\rm eff},0}}
{G_{{\rm eff},0}}$. The best present bounds are:
\bea
\gamma_{PN}-1 &=& (2.1\pm 2.3)\cdot 10^{-5}\nonumber\\
\beta_{PN}-1 &=& (0\pm 1)\cdot 10^{-4} \nonumber\\
\frac{{\dot G}_{{\rm eff},0}}{G_{{\rm eff},0}} &=& (-0.2\pm 0.5)
\cdot 10^{-13}~y^{-1}~. \lb{limit}
\eea
where the first bound was
obtained from the Cassini mission \cite{BIT03} and the other two
from high precision ephemerides of planets \cite{Pit05}~(the
second bound has been recently confirmed by the Lunar Laser
ranging \cite{WTB05} -- their value is $\beta_{PN}-1 = (1.2\pm
1.1)\cdot 10^{-4}$).

As a consequence of the smallness of $\gamma_{PN}-1$, the
Brans-Dicke parameter $\omega_{BD}$ satisfies {\it today} the
inequality
\be
\omega_{BD,0} > 4 \times 10^4~. \lb{BD}
\ee
The resulting bound on $F'(0)$ is very stringent
\be
\frac{F'(0)}{\sqrt{F_0}} = \pm
\frac{\Phi'(0)}{\sqrt{\omega_{BD,0}}} ~.\lb{dF0}
\ee
However, as was discussed in \cite{BEPS00}, the quantity
$\omega_{BD}$ need not be so large as (\ref{BD}) in the past,
though one can deduce a looser inequality applying up to redshift
$z\lesssim 1$ with fairly reasonable assumptions. From (\ref{Phi},
\ref{dF0}) we can derive the allowed range of initial values
$F'(0)$ and we find $|F'(0)|\sim \omega_{BD,0}^{-\frac{1}{2}}$ and
$\Omega_{DE,0}-\Omega_{U,0}>0$, the result that we will recover in
Sec. 5 when performing the small $z$ expansion of all quantities.
The peculiar case $F'(0)=0$ together with
$\Omega_{DE,0}-\Omega_{U,0}>0$ corresponds to pure General
Relativity today ($\omega_{BD}=\infty$).

As noted above, supernova observations permit DE to be of the
phantom type, with the equation of state parameter $w_{DE}<-1$ at
small redshifts. This is a strong motivation for considering DE
models in the framework of the scalar-tensor theory of gravity. More
generally, at present there is much interest in models with modified
gravity of which the scalar-tensor theory is a well known
representative. In scalar-tensor DE models, a meaningful definition
of energy density and pressure of the DE sector requires some care
(see also \cite{T02} for a detailed explanation). Let us {\it define}
the energy density $\rho_{DE}$ and the pressure $p_{DE}$ in the
following way:
\bea
3F_0~H^2 &=& \rho_m + \rho_{DE} \lb{E1a} \\
-2F_0~{\dot H} &=&  \rho_m + \rho_{DE} + p_{DE} ~. \lb{E2a}
\eea
This just corresponds to the (conventional) representation of the
true equation for scalar-tensor gravity interacting with matter
in the {\em Einsteinian} form with the constant $G_0=G_N(t_0)$:
\be
R_{\mu\nu}- {1\over 2} R\,g_{\mu\nu}=8\pi G_0\left(T_{\mu\nu,m}+
T_{\mu\nu,DE}\right)~. \lb{E3a}
\ee
With these definitions, the usual conservation equation applies:
\be
{\dot \rho_{DE}} = -3H ( \rho_{DE} + p_{DE} )~.
\ee

If we define the equation of state parameter $w_{DE}$ through
\be
w_{DE} \equiv \frac{p_{DE}}{\rho_{DE}}~,
\ee
the time evolution of
the DE sector is given by
\be
\frac{\rho_{DE}(z)}{\rho_{DE,0}}
\equiv \epsilon (z) = \exp \left[ 3\int_{0}^z
dz'~\frac{1+w(z')}{1+z'}\right]~.\lb{gz}
\ee
Using (\ref{E1a},\ref{E2a}), one gets
\be
w_{DE} = \frac{\frac{2}{3}(1+z)\frac{d\ln H}{dz} - 1}{1 -
\frac{H_0^2}{H^2}~\Omega_{m,0}(1+z)^3}~,\lb{wDE}
\ee
where
\be
\Omega_m \equiv \frac{\rho_m}{3H^2 F_0}~.\lb{Om}
\ee
Then Eq.(\ref{E1a}) can be rewritten as
\be h^2(z)= \Omega_{m,0} ~(1+z)^3
+ \Omega_{DE,0} ~\epsilon(z)~\lb{hz}
\ee
where $\Omega_{DE,0}= 1 -\Omega_{m,0}$ by definition. The
condition for DE to be of the
phantom type, $w_{DE}<-1$, can be obtained from (\ref{wDE}). It
reads \cite{SS00,BEPS00} \be \frac{d h^2}{dz} < 3 ~\Omega_m
~(1+z)^2~.\lb{ineqP} \ee This inequality is modified in the
presence of spatial curvature \cite{PR05}. As was first emphasized
in \cite{BEPS00}, the scalar-tensor gravity allow for phantom DE.
Indeed, for these models
\be
\rho_{DE} + p_{DE} = \dot \Phi^2 +
{\ddot F} - H {\dot F} + 2(F - F_0)~{\dot H}~,\lb{wec}
\ee
hence the weak energy condition for DE can be violated (see also
\cite{T02}). Moreover, as it will be shown below, the weak energy
condition may be violated even for the sum of DE and
non-relativistic matter, i.e. for the whole right-hand side of
Eqs. (\ref{E3a}), leading to $dh/dz < 0$. However, such a strong
violation corresponding to $w_0\equiv w_{DE}(0) <
(1-\Omega_{m,0})^{-1}\approx -1.4$ is not supported by the
existing data (though not completely excluded either).

The relation between the Hubble parameter and the luminosity
distance in scalar-tensor gravity is the same as in GR
\footnote{Note that the definitions of $d_L(z)$ and $D_L(z)$ are
interchanged in \cite{CP01} and \cite{PR05}. }:
\be h^{-1}(z)=
\left( \frac{d_L(z)}{1+z}\right)'~,~~ d_L(z)=H_0~D_L(z)~. \lb{hd}
\ee
However, as discussed in \cite{GGI99,RU02}, when using
supernova data to obtain $D_L(z)$, one has to take into account
the dependence of the Chandrasekhar mass on $G_{\rm eff}$, so that
the SnIa peak luminosity appears to be $\propto (G_{\rm
eff}(z)/G_{\rm eff,0})^{-3/2}$. As shown in \cite{BEPS00}, the
full reconstruction of the functions $F(\Phi),~U(\Phi)$ requires
two independent types of observations: $d_L(z)$ or another
function of $z$ that probes the background evolution, and the
growth factor of matter perturbations $\delta \rho_m(z)/\rho_m(z)$
at some comoving scale much less than the Hubble scale, too. On
the other hand, as emphasized in \cite{EP01}, one can already
obtain powerful constraints from a partial reconstruction using
$D_L(z)$ only. Such a partial reconstruction is possible when some
additional condition is imposed on either $F$ or $U$. This is the
way we adopt in Sec.~7: we reconstruct the function $F$ for a
given Hubble parameter $H(z)$ and a constant potential $U$ and
investigate whether the resulting model is viable. Before
embarking on such partial reconstructions, we derive first an
integral form of the general solution of Eq. (\ref{MF}).

\section{The master equation for $F$}
Let us now consider the master equation (\ref{MF}) and present its
general solution in an integral form. This will allow us to understand
general properties of solutions, in particular, their dependence on
initial conditions. The first step is to note that one solution of the
homogeneous equation (\ref{MF}) (without source term) is given by
\be
F \propto (1+z)^2~.\lb{x2}
\ee
This suggests us, following \cite{EP01}, to introduce the function $f$
defined as follows
\be
\frac{F(z)}{F_0} \equiv (1+z)^2 f(1+z)~,\lb{f}
\ee
in terms of which Eq.(\ref{MF}) becomes
\be
f'' + (\ln h)' f' = \frac{6~u}{x^4 ~h^2}~\Omega_{U,0} + \frac{3}{x ~h^2}
~\Omega_{m,0}~,\lb{Mf}
\ee
where we have introduced the variable $x\equiv 1+z$. Due to the absence
of any term proportional to $f$, in accordance with (\ref{x2},\ref{f}),
Eq. (\ref{Mf}) is easily integrated formally, and its general solution
has the integral form
\bea
f(x) = 1 + \left[ -2 + \frac{F'(x=1)}{F_0} \right]~\frac{d_L}{x} &+
& 6~\Omega_{U,0}~\int_1^x \frac{dx'}{h(x')}~
\int_1^{x'} \frac{u(x'')~dx''}{x''^4~h(x'')} \nonumber \\
&+& 3~\Omega_{m,0}~\int_1^x \frac{dx'}{h(x')}~
                     \int_1^{x'} \frac{dx''}{x''~h(x'')}~.\lb{fsol}
\eea

A nontrivial dependence on initial conditions is through the second
term only. We see that for a given initial value $F'(z=0)$, this term
is proportional to the dimensionless luminosity distance $d_L$. Due
to Eq. (\ref{dF0}), we have $f'(x=1)= -2 \pm \frac{|\Phi'(z=0)|}
{\sqrt{F_0~\omega_{BD,0}}}\approx -2$, indicating that we must start
today, on observational grounds, very close to GR. Hence, the second
term of the general solution is bound to be negative and it is this
term that will possibly push the quantity $f(z)$ downwards for
increasing $z$. Finally, the corresponding quantity $F$ is trivially
obtained using (\ref{f}). If we {\it choose} $U$ to be constant, we
can implement the reconstruction of the microscopic Lagrangian. By
inspection of (\ref{MF}) for constant $U$, we find the asymptotic
solution
\be
F(x) = C_1~x^2 + C_2~x^{\frac{3}{2}} + F_0~,\lb{assol}
\ee
in complete agreement with (\ref{fsol}) setting $u=1$. We can now
proceed with the general reconstruction scheme \cite{BEPS00,EP01}
and find $\Phi(z)$, and hence $z(\Phi)$, by integration of
(\ref{Phi}). This would finally give us $F(\Phi)$.

Let us come back to the solution (\ref{x2}). It corresponds to the
parametrization $Z=-1$ with
\be
F(\Phi) = \frac{1}{6} \Phi^2~,~~\omega_{BD} = -\frac{3}{2}~,
\lb{cc}
\ee
where $\Phi$ is defined up to some constant. From (\ref{cc}) it is
seen that this solution is unphysical in view of (\ref{posEn}), see
\cite{EP01}. It is interesting that (\ref{cc}) corresponds
to a conformally coupled scalar field in the Jordan frame satisfying
\be
{\ddot \Phi} + 3 H {\dot \Phi} + \frac{R}{6} \Phi = 0 \lb{cc1}~.
\ee

\section{Scaling solutions}
Let us now consider the so-called scaling solutions for which DE
scales as some power of $a$,
\be
\rho_{DE}\propto a^{-3\gamma}~,
\ee
which attracted a lot of interest previously.\footnote{Our
definition for $\rho_{DE}$, Eq.(\ref{E1a}), differs from that used in
\cite{PBP05}, and this explains why our results differ from theirs.}
Clearly, for these solutions DE has an effective barotropic (constant)
equation of state, viz.
\be
w_{DE} = -1 + \gamma~,
\ee
and we can readily write formally the general scaling solution by
substituting
\be
h^2 = \Omega_{m,0}~(1+z)^3 + \Omega_{DE,0}~(1+z)^{3\gamma} ~,\lb{hn}
\ee
into the integral expression (\ref{fsol}). We emphasize that this
gives the general scaling solution irrespective of any limiting case
and for an arbitrary potential, the only assumption is that of the
spatial flatness. Using (\ref{fsol}) and (\ref{hn}), general analytic
expressions for $F(z)$ can be obtained only for the cases $\gamma =
1,~4/3$ (see Appendix).

In analogy with a minimally coupled scalar field (quintessence), a
scaling solution satisfies
\be
\frac{\rho_{DE} + p_{DE}}{\rho_{DE}} = 1 + w_{DE} = \gamma~.\lb{n/3}
\ee
However, in contrast to the minimally coupled scalar field case with
a positive potential, (\ref{n/3}) does not imply $0\le \gamma \le 2$,
in particular $\gamma$ can be negative which corresponds to phantom
DE.

To get insight into the ability of scalar-tensor DE models to produce
various equation of state parameters $w_{DE}$, it is instructive to
study first scaling solutions in the absence of dust-like matter
($\rho_m=0$). In this way, a lower limit on $w_{DE}$ for realistic
solutions with $\rho_m\not= 0$ can be obtained. These scaling
solutions can also be considered as describing the asymptotic future
of our universe when $\Omega_m\to 0$ and $\Omega_{DE}\to 1$. Then $a(t)
\propto |t|^q$ with $q={2\over 3\gamma}$. For phantom (or,
super-inflationary) solutions, $q<0$ and then the moment $t=0$
corresponds to the 'Big Rip' singularity \cite{Cald02,St00}.

It is straightforward to check that such scaling solutions can exist
only when $F=\alpha \Phi^2$ with $\alpha = const$. This form of $F$
corresponds to a constant Brans-Dicke parameter $\omega_{BD}=
\frac{1}{4\alpha}$ for $Z(\Phi)=1$. Further on, we assume that
$\alpha > 0$ for stability of the theory. The two scaling solutions in
the vanishing potential case (the pure Brans-Dicke theory) are
\cite{HT72}
\be
|\Phi|\propto |t|^r,~~~r={1\pm 3\sqrt{1+{1\over 6\alpha}}
\over 8\left(1+{3\over 16\alpha}\right)}~,~~~ q={ 1+{1\over 4\alpha}
\mp \sqrt{1+{1\over 6\alpha}} \over 4 \left(1+{3\over 16\alpha}\right)}
> 0~.\lb{ssU0}
\ee
%
%
Hence it is not possible to get a scaling solution with $w_{DE}<-1$
with vanishing (or negligible) potential $U(\Phi)$. We note that the
above solution reduces to (\ref{x2},\ref{cc}) obtained for $Z=-1$ when
we formally put $\alpha=-\frac{1}{6}$ in (\ref{ssU0}). Indeed we get
$q=1,~r=-1$, hence $F\propto a^{-2}$ which is exactly the solution
(\ref{x2}).

However, scaling solutions supported by both the kinetic and the
potential energy of $\Phi$ can exist in the presence of a polynomial
potential $U(\Phi)=U_0|\Phi|^n$ (we assume that the potential is
positive, so $U_0>0$). We have for these solutions
\be
r={2\over 2-n}~,~~~q={2\left(n+2+{1\over \alpha}\right)\over (n-2)(n-4)}
\label{scale2}
\ee
while the inequality $q^2+2qr > \frac{r^2}{6\alpha}$ has to be
satisfied (this is always the case if $\alpha \ll 1$ and
$n={\cal O}(1)$). They were first found in \cite{BM90} (see also the
analysis of their stability in \cite{A99,HW00}) but they have not been
discussed yet in connection with phantom DE. Note that there is no
phantom behaviour for usually considered quadratic and quartic
potentials.

So, we see that it is possible to get scaling solutions with $q<0,
~w_{DE}<-1$ if $2<n<4$. For these solutions, $r<0$, too, so that $a(t)$
diverges in the "Big Rip" singularity at some finite moment of time in
future. It is clear that, relaxing the requirement of scaling
behaviour, it is possible to add some amount of dust-like matter to
these solutions while still keeping the phantom behaviour of DE. However,
the amount of 'fantomness' exhibited by them, i.e. the modulus of the
minimal possible value of $w_{DE} +1=2/3q$, is very small
for small values of $\alpha$ (equivalently large $\omega_{BD}$), viz.
\be
w_{DE} +1 \ge -{\alpha \over 3}~,
\ee
where the equality is achieved for $n=3$. Thus, the conclusion is
that polynomial potentials and scaling solutions in viable
scalar-tensor DE models can lead to violation of the WEC, however,
by the small amount $\sim\omega_{BD}^{-1}$ only.

\section{Small $z$ expansion and Solar System gravity tests}
Of course, scaling solutions considered in the previous section are
very specific ones. Let us now study generic solutions describing DE
in the scalar-tensor gravity. Since, as was explained in the
Introduction, if DE crosses the phantom boundary at all, it has done
it in a very recent epoch, at small $z$, it is natural to study the
expansion of a generic solution in powers of redshift $z$. For each
solution $H(z),~\Phi(z)$, the basic microscopic functions $F(\Phi)$
and $U(\Phi)$ can be expressed as functions of $z$ and expanded into
Taylor series in $z$:
\bea
\frac{F(z)}{F_0} = 1 + F_1 ~z + F_2 ~z^2 + ...> 0~,\lb{expz1}\\
\frac{U(z)}{3F_0~H_0^2}\equiv \Omega_{U,0}u = \Omega_{U,0} + u_1 ~z
+ u_2 ~z^2 + ...~.
\lb{expz2}
\eea
Note that this expansion produces two parameters in each order of $z$
which are independent of initial conditions and can be expressed
through derivatives of $F(\Phi)$ and $U(\Phi)$ with respect to $\Phi$.
The corresponding expansion for $\Phi'(z)$ is:
\bea
(F_0)^{-\frac{1}{2}}\Phi'(z) &=& \Phi'_0 + \Phi'_1~z + \Phi'_2~z^2
+ ... \nonumber\\
&=& \Delta + \frac{1}{\Delta}\Bigg[6(F_1-F_2+\Omega_{U,0}-1)-
3(\Omega_{m,0}+u_{1}) \nonumber\\
&+&\frac{3(\Omega_{m,0}+\Omega_{U,0})}{\frac{F_{1}}{2}-1}\,(4F_{1}
-2F_{2}+6\Omega_{U,0}+3\Omega_{m,0}-6)\Bigg]\cdot z+\dots \lb{expPhi}
\eea
with $\Delta^2\equiv 6~(\Omega_{DE,0}-\Omega_{U,0}-F_{1})$. As we will
see below, $\Delta^2>0$. In principle, the expansion (\ref{expPhi})
can be inverted to get $z(\Phi)$. From (\ref{expz1},\ref{expz2}), all
other expansions can be derived:
\bea
h^2(z) &=& 1 + h_1 ~z + h_2 ~z^2 + ...~,\\
\epsilon(z) &=& 1 + \epsilon_1 ~z + \epsilon_2~z^2 + ...> 0~,
\lb{expgz}\\
w_{DE}(z) &=& w_0 + w_1 ~z + w_2 ~z^2 + ...~,\\
H_0^{-1}\frac{ {\dot G}_{\rm eff} }{G_{\rm eff}} &=& g_0 + g_1~z +
g_2~z^2 + ....,
\eea
which can be used in order to constrain parameters of our model.

There are two types of observational constraints at small
redshifts. The first of them includes those ones that follow from
Solar System and other tests of possible deviations of the
scalar-tensor gravity from GR at the present moment ($z=0$); in
particular, $|g_0|\lesssim 10^{-3}$ at the $2\sigma$ confidence
level from the last of Eqs. (\ref{limit}). Other constraints
follow from cosmological tests and refer to the whole range of
redshifts up to $z\sim 1$ and higher redshifts, depending on the
nature of the
test. In particular, if we assume that the present supernova data
admit (or even favour) $w_{DE}(z) < -1$ for $z\lesssim 0.3$ as was
argued in the Introduction, then ${d\epsilon\over dz}=\epsilon_1 +
2~\epsilon_2~z + 3~\epsilon_3~z^2 + ....< 0$ for $z\lesssim 0.3$.
In particular, we must have $\epsilon_1 < 0$ that is equivalent to
Eq. (\ref{ineqP}) taken at $z=0$.

It follows from the substitution of these expansions into Eqs.
(\ref{MF},\ref{Phi}) that
\bea
h_{1}&=&\frac{1}{1-\frac{F_{1}}{2}}\big(6-3\Omega_{m,0}-6\Omega_{U,0}
-4F_{1}+2F_{2}\big)~,\lb{h1} \nonumber\\
h_{2} &=& \frac{3}{(\frac{F_{1}}{2}-1)^2}\Bigg[ F_1\left(
\frac{5}{2}F_1-3 F_{2}- \frac{F_3}{2} + 4\Omega_{U,0} + \frac{u_1}{2}
+\frac{11}{4}\Omega_{m,0} - 7 \right)\nonumber \\
 &&   +F_{2}^2-3F_{2}\Omega_{U,0}-\frac{3}{2}F_{2}\Omega_{m,0}+4F_{2}
 + F_{3}-5\Omega_{U,0}-u_{1}-4\Omega_{m,0}+5\Bigg]~,\\
1+ w_0 &=&\frac{\frac{5}{2}F_{1}-2F_{2} - 6( \Omega_{DE,0} -
\Omega_{U,0} )+ \frac{3}{2}F_1~\Omega_{DE,0}} {3\Omega_{DE,0}
(\frac{F_1}{2}-1)}~, \lb{cp0}\\
w_{1} &=& \frac{1}{3\Omega_{DE,0}}\Bigg[ \frac{1+6\frac{\Omega_{m0}}
{\Omega_{DE,0}}}{\frac{F_{1}}{2}-1}\bigg(4F_{1}-2F_{2}+6\Omega_{U,0}-
3\Omega_{DE,0}-3\bigg)-9\frac{\Omega_{m0}}{\Omega_{DE,0}} \nonumber \\
&+& \frac{6}{(\frac{F_{1}}{2}-1)^2}\bigg( \frac{5}{2}F_{1}^2-3F_{1}
F_{2}-\frac{F_{1}F_{3}}{2}+4F_{1}\Omega_{U,0}+\frac{F_{1}u_{1}}{2} +
\frac{11}{4}F_1 \Omega_{m0}-7F_1\nonumber \\
&+& F_2^2-3F_{2}\Omega_{U,0}-\frac{3}{2}F_{2}\Omega_{m0} + 4F_{2}+
F_{3}-5\Omega_{U,0}-u_{1}+4\Omega_{DE,0}+1\bigg)\Bigg]\nonumber \\
&-& \frac{1}{3\Omega_{DE,0}^2(\frac{F_{1}}{2}-1)^2}\bigg(4F_{1}-
2F_{2}+6\Omega_{U,0}-3\Omega_{DE,0}-3\bigg)^2~.
\eea
The quantities $F_1,~F_2,~\Omega_{DE,0} - \Omega_{U,0}$ satisfy
important constraints. For $\omega_{BD,0}$, we have the expression
\be
\omega_{BD,0} = \frac{6(\Omega_{DE,0} - \Omega_{U,0} - F_1)}{F_1^2}
= \frac{\Delta^2}{F_1^2} \lb{BD0}
\ee
which should be very large and positive, see (\ref{BD}). Therefore, 
we must have $|F_1|\ll 1$ and $\Delta^2\approx
6(\Omega_{DE,0} - \Omega_{U,0})>0$.\footnote{Note that this condition
is not satisfied in the recent paper \cite{NP06} that results in
unphysical nature of its best-fit solution for $z<0.2$.} Moreover,
since $\Delta^2 < 6 \Omega_{DE,0}< 5$ for positive $U$,
\be
|F_1|<\left( {5\over \omega_{BD,0}}\right)^{1/2}\lesssim 10^{-2}~.
\lb{F1}
\ee

Thus, two cases are possible. In the first case, the further
coefficients $F_2,F_3$, etc. in the expansion (\ref{expz1}) are all
of the order of $F_1$, i.e. they satisfy the inequality (\ref{F1}),
too. In this case, the first derivative of $F$ with respect to $z$
or $\Phi$ at the present moment is not atypical compared to other
derivatives. Then, however, a possible amount of phantomness
(the quantity defined in the end of the previous section) is also
of the order of $|F_1|$, i.e. less than $1\%$. Such DE will be
practically indistinguishable from a cosmological constant.

Another possibility which admits 'significant phantomness', namely,
$\min(1+w_{DE}(z))< - 0.01$, takes place if $|F_2|,~|F_3|,~
(\Omega_{DE,0}- \Omega_{U,0})$ and so on are significantly larger than
$|F_1|$. It is a matter of taste if one considers the second
possibility as 'fine tuned' with respect to the first one;
observations should finally tell us if significant
phantomness does exist or not. In any case, it is clear that if we are
interested in any prediction for a significant deviation from the
cosmological constant variant of DE, we may neglect $F_1$ as compared
to $F_2$ and other parameters (but not in those expressions where it
enters as a multiplier).

For $|F_1|\ll 1$, all the expansions above simplify significantly,
and we have, in particular,
\bea
1 + w_0 &\simeq&\frac{2F_{2} + 6(\Omega_{DE,0} - \Omega_{U,0})}
{3\Omega_{DE,0}}~. \lb{cp0a}
\eea
From (\ref{cp0a}), the necessary condition to have phantom DE {\it
today} reads
\be \left({d^2F\over d\Phi^2}\right)_0={F_2 \over
3~(\Omega_{DE,0} - \Omega_{U,0})} < -1 ~. \lb{cp2}
\ee
In particular, $F_2<0$, because $\Omega_{DE,0} - \Omega_{U,0}>0$ as
discussed above.\footnote{In this place our considerations
intersect with those of the recent paper \cite{MSU06}. The
condition (\ref{cp2}) corresponds to $\beta_0>1$ in the notation
of that paper. The other case mentioned there, $\beta_0\sim -1$,
does not lead to significant fantomness of DE with our definition
of the DE energy-momentum tensor.}
In the same limit, $h_1<0$, i.e. the WEC is violated for the total
effective energy-momentum tensor for matter + DE (for the whole
right-hand side of Eq. (\ref{E3a})), if the stronger inequality is
fulfilled:
\be
F_2 < - {3\over 2}(1+\Omega_{DE,0}-2\Omega_{U,0})~. \lb{h1neg}
\ee
The expression for $w_1$ becomes in the limit $|F_1|\ll 1$
\bea
w_1 &=& \frac{1}{3 \Omega_{DE,0}}\left[\left(1+6\frac{\Omega_{m,0}}
              {\Omega_{DE,0}}\right)(2F_2+3+3\Omega_{DE,0}-6\Omega_{U,0})-
                                           9\frac{\Omega_{m,0}}{\Omega_{DE,0}}\right]\nonumber \\
&+& \frac{2}{\Omega_{DE,0}}\left[F^2 \left(F_2 - 3\Omega_{U,0}-
       \frac{3}{2}\Omega_{m,0}+4\right) - 5\Omega_{U,0}+4\Omega_{DE,0}+1 +F_3 - u_1\right]\nonumber \\
&-& \frac{1}{3\Omega^2_{DE,0}}\left(2F_2+3\Omega_{DE,0}+3-6\Omega_{U,0}\right)^2 \lb{w1a} \\
&=& \frac{9(w_0-1)^2\Omega_{DE,0}-8 - 6w_0^2 + 9\Omega_{m,0} - 18\Omega_{U,0} +
    w_0\left(26 - 9\Omega_{m,0} + 18\Omega_{U,0}\right)}{2}\nonumber \\
&+&  \frac{ \left( - 6w_0 + 14\Omega_{U,0} + 3\Omega_{m,0}(-3\Omega_{U,0} + 2w_0) \right)}{\Omega_{DE,0}}
        +  \frac{2}{\Omega_{DE,0}} ~(F_3 - u_1)\lb{w1b} ~,
\eea
where the last equality (\ref{w1b}) is obtained using (\ref{cp0a}). Since $F_3$ and $u_1$ are
free parameters determined by a concrete choice of $F(\Phi)$ and $U(\Phi)$, it is well
possible to have $w_1>0$ for $w_0 < -1$ in order to realize a smooth phantom divide crossing
at small $z$ for DE in the scalar-tensor gravity.

We obtain further
\bea
g_0
%
%
&=& F_{1}\Bigg\{ 1-\frac{1}{(\frac{F_{1}}{2}-1)(F_{1}^2-3F_{1}-
3\Omega_{U,0}+3\Omega_{DE,0})(F_{1}^2-4F_{1}-4\Omega_{U,0}+
4\Omega_{DE,0})}\nonumber \\
&& \quad \times\Bigg[\bigg((F_{1}^2-4F_{2})(\frac{F_{1}}{2}-1)
-(5F_{1}-2F_{2}+6\Omega_{U,0}-3\Omega_{DE,0}-5)F_{1}\bigg)\nonumber \\
&&\qquad \times(F_{1}+\Omega_{U,0}-\Omega_{DE,0})+(4F_{1}-2F_{2}+
6\Omega_{U,0}-3\Omega_{DE,0}-3)F_{1}^2\nonumber\\
&&\qquad  +(F_{1}F_{2}-\frac{5}{2}F_{1}+\frac{F_{1}u_{1}}{2}-
\frac{3}{2}F_{1}\Omega_{DE,0}-6\Omega_{U,0}-u_{1}+6\Omega_{DE,0})
F_{1}\bigg]\Bigg\}~. \lb{dG0a}
\eea
Expression (\ref{dG0a}) simplifies considerably for $|F_1|\ll 1$:
\be
g_0\simeq F_1 \left( 1-\frac{ F_2 }{ 3(\Omega_{DE,0} -
\Omega_{U,0}) } \right) = F_1\left(1-\left({d^2F\over d\Phi^2}
\right)_0\right)~.\lb{dG0b}
\ee
Note that $g_0$ and $F_1$ have the same sign for the case of
(significantly) fantom DE. Finally, for the post-Newtonian parameter
$\beta_{PN}$ and $\gamma_{PN}$ we have
\bea
\beta_{PN} &=& 1 + \frac{\omega_{BD}'}{4(2\omega_{BD}+3)
(\omega_{BD}+2)^2}\frac{F}{F'} \lb{beta}\\
\gamma_{PN} &=& 1 - \frac{1}{\omega_{BD} + 2}~.\lb{gamma}
\eea
Taking the zeroth order in $z$, we obtain :
\bea
\beta_{PN,0} &=& 1 + \frac{\omega_{BD,1}}{4(2\omega_{BD,0}+3)
(\omega_{BD,0}+2)^2}\frac{1}{F_{1}}~.\lb{beta0}\\
\gamma_{PN,0} &=& 1 - \frac{1}{\omega_{BD,0} + 2}~.\lb{gamma0}
\eea
 For $|F_1|\ll 1$ and $|F_1u_1/\Delta^2|\ll 1$, we have:
\bea
\omega_{BD,0} &\simeq& 6(\Omega_{DE,0}-\Omega_{U,0})~\frac{1}
{F_1^2}\\
\omega_{BD,1} &\simeq& 24(\Omega_{DE,0}-\Omega_{U,0})~\frac{F_2}
{F_1^3}~,
\eea
which finally yields
\be
\gamma_{PN,0} = 1 - {F_1^2\over 6(\Omega_{DE,0} - \Omega_{U,0})}~,
~~~\beta_{PN,0} = 1 - \frac{F_1^2 ~F_2}{72(\Omega_{DE,0} -
\Omega_{U,0})^2}~.\lb{beta0a}
\ee

Now we can use (\ref{cp0},\ref{BD0},\ref{dG0b},\ref{beta0a})
to extract information from the Solar System
constraints. First we note that all the Solar System constraints
(\ref{BD}),(\ref{limit}) are satisfied for a sufficiently small
$|F_1|$ (less than the upper bound (\ref{F1})) independently of any
requirement concerning the present DE equation of state, in
particular, whether one has today DE of the phantom type or not.
From (\ref{gamma}), it is seen that $\gamma_{PN,0}\approx 1$ and
$\gamma_{PN} < 1$ for any viable scalar-tensor model of DE having
$\omega_{BD,0}$ positive and large. For significantly phantom DE,
other constraints follow from (\ref{beta0a}):
\bea
\beta_{PN,0} &>& 1~, \lb{cb} \\
\frac{\gamma_{PN,0} - 1}{\beta_{PN,0} - 1} = \frac{12~(\Omega_{DE,0}
 - \Omega_{U,0})}{F_2} &=& 6~\Omega_{DE,0}~\frac{1 + w_0 }
 {F_2} - 4 ~.\lb{b/g2}
\eea
Therefore, for significantly phantom DE
\be
-4 < \frac{\gamma_{PN,0} - 1}{\beta_{PN,0} - 1} < 0~.
\ee

It is possible to invert formulas (\ref{dG0b},\ref{beta0a}) and
express $F_1,~F_2$ and $\Omega_{DE,0}-\Omega_{U,0}$ through the
Solar System observables $\gamma_{PN,0},~\beta_{PN,0}$ and $g_0$:
\bea
F_1 &=& g_0~\frac{\gamma-1}{\gamma-1 - 4(\beta-1)}\\
F_2 &=& -2~g_0^2~ \frac{\beta-1}{[ \gamma-1 - 4(\beta-1) ]^2}\\
\Omega_{DE,0}-\Omega_{U,0} &=& -\frac{1}{6}~g_0^2~ \frac{\gamma-1}
                    {[ \gamma-1 - 4(\beta-1) ]^2}\\
1+w_0 &=& -\frac{1}{3}~g_0^2~\frac{4(\beta-1) + \gamma-1}{\Omega_{DE,0}~[ \gamma-1 - 4(\beta-1) ]^2}
\eea
Thus, in principle, it is possible to test $w_0 <-1$ in the Solar
System as was recently discussed in \cite{MSU06}. However, this
may be very difficult to perform in practice since the small
parameter $F_1$ enters quadratically into $\gamma_{PN,0} - 1$ and
$\beta_{PN,0} - 1$ while not appearing (in some limit) in
$1+w_0$. So, a rather significant fantomness of DE is typically
accompanied by very small deviations of the post-Newtonian
parameters from their GR values. E.g., let us take
$\Omega_{DE,0}=0.7,~\Omega_{U,0}=0.6,~\gamma_{PN,0}-1= -2\cdot
10^{-5}$ (the latter being marginally possible at the $2\sigma$
level) and $w_0 = -1.2$. Then, from Eqs.
(\ref{beta0a},\ref{cp0a}), we get $|F_1|=3.5\cdot 10^{-3},~
F_2=-0.51$ and $\beta_{PN,0}-1=0.85\cdot 10^{-5}$ -- an order of
magnitude below the present upper limit.\footnote{Let us emphasize
once more that the formulas (\ref{dG0b},\ref{beta0a}) were
obtained under the assumption $|F_2|\gg |F_1|$. For this reason,
they are not applicable, e.g., to the scaling solution
(\ref{scale2}) for which $F_2=-F_1/2 >0,~\gamma_{PN}=1-4\alpha$ in
the limit $\alpha \ll 1$, and $\beta_{PN}\equiv 1$.} Further,
$w_1=2.9\,(F_3-u_1)- 0.70$. Finally, in this case $|g_0|=0.95\cdot
10^{-2}$ that is an order of magnitude larger than the $2\sigma$
upper limit following from the last of Eqs.~(\ref{limit})~! So, if
this upper bound will be confirmed, $|F_1|$ has to be decreased by an
order of magnitude which results in $\beta_{PN,0}-1$ and
$\gamma_{PN,0}-1$ being on the level of $10^{-7}$. For comparison,
in the extreme opposite case $\Omega_{U,0}=0$ with the same values
of $\Omega_{DE,0},~ \gamma_{PN,0}$ and $w_0$, we get
$|F_1|=9.2\cdot 10^{-3},~ F_2=-2.3,~\beta_{PN,0}-1= 0.55\cdot
10^{-5},~w_1=2.9\,(F_3-u_1)+1.5$ and even larger $|g_0|=1.9\cdot
10^{-2}$.

This shows also that the measurement of $\dot G_{{\rm eff},0}/
G_{{\rm eff},0}$ is the most critical among Solar System tests of
scalar-tensor DE since this quantity is proportional to the first
power of the small parameter $|F_1|$ (apart from the exceptional case
$(d^2F/d\Phi^2)_0=1$ which does not lead to the present phantom
behaviour of DE). Also, to determine $w_1$ which is necessary
in order to consider the possibility of phantom boundary crossing,
the determination of $(d^2\ln G_{\rm eff}/dt^2)_0$ is required, something
that is hardly possible. Thus, testing the phantom behaviour of
scalar-tensor DE in the Solar System may be much more difficult
than in cosmology.

\section{General reconstruction of $F(z)$}
We consider now the reconstruction of $F(z)$ for given Hubble parameter $H(z)$ and potential $U$.
As can be seen from (\ref{hz},\ref{gz}), $H(z)$ is a functional of $w(z)$. We will consider
several cases corresponding to phantom divide crossing at very small redshifts as favoured
by the latest observations. It is desirable to derive some general properties of the behaviour
of our system before embarking on the study of specific models.

It is easy to derive the following general property shared by a large class of models: if we
require that $\phi'^2>0$ and also $U\ge 0$, then the following inequality must be satisfied
\be
\frac{F'}{F_0}\left ( \frac{(1+z)^2}{4}\frac{F'}{F} - (1+z)\right ) + \frac{F}{F_0} >
                   (1+z)^3 h^{-2}(z)\Omega_{m,0}~.\lb{ineq}
\ee
Note that the r.h.s. of (\ref{ineq}) tends to one in any model for which $w_{DE}<0$ and it will tend
to this asymptotic value quickly for $w_{DE}<-0.5$. We can consider several particular cases:
\begin{itemize}
\item
It is seen from (\ref{ineq}) that models which at some redshift $z_m\gg 1$ satisfy $F'(z_m)=0$
and $0<F(z_m)<1$ sufficiently small, must also have $\phi'^2(z_m)<0$, such an example is actually
illustrated in Figure 1.
\item
When $F\to 0$ for $z\to z_m$ while at the same time $F'(z_m)\ne 0$, it follows from the expression
(\ref{posEn1}) for $\phi'^2$ and using (\ref{Phi}) that $\phi'^2\to \infty$.
\item
Finally, an interesting case is provided when $F(z_m)=F'(z_m)=0$ for some $z_m$. We see first from
(\ref{Phi}) that $\Phi'^2\to \Phi'^2(z_m)$, where  $\Phi'^2(z_m)$ is a (small) negative number.
Let us assume that $\frac{F'}{F}$ is bounded when $z\to z_m$. In that case the inequality (\ref{ineq})
cannot be satisfied, hence $\phi'^2(z_m)<0$ (allways assuming $U>0$) and actually we have
$\phi'^2\to -\infty$ for $z\to z_m$ as can be checked directly with the definition of $\phi'^2$.
As a consequence, if $\phi'^2>0$ as $z\to z_m$ then we must have $\left|\frac{F'}{F}\right|\to \infty$
and if in addition $\frac{F'^2}{F}$ is bounded, then $\phi'^2\to \infty$.
\end{itemize}
Some of these properties are illustrated with Figures 1,2,3.

Let us consider now the space-time background evolution encoded in the quantity $h(z)$.
From a theoretical point of view, as is seen from (\ref{hz}), a given function $w(z)$
implies a corresponding  functional form $h(z)$, and conversely using (\ref{wDE}).
Future data are expected to mesure $D_L(z)=H_0^{-1}d_L(z)$, and therefore $h(z)=H_0^{-1}~H(z)$,
with high precision. In the meantime, we can try some particular expressions $w(z)$, or $h(z)$,
parametrized with the help of a limited number of free parameters, an attitude which
turns out to be very fruitful.
We will consider both constant and variable equation of state parameter $w$. Observations
suggest that $w$ can be varying and we will model this variation using the following
two-dimensional parametrization of the equation of state parameter suggested in
\cite{CP01}, \cite{L03})
\be
w(z) =  (-1 + \alpha) + \beta ~(1-x) \equiv w_0 + w_1~\frac{z}{1+z}~.\lb{wCPL}
\ee
where $x\equiv \frac{a}{a_0}$. We will sometimes compare our results with the parametrization
suggested in \cite{SSSA03}
\be
\epsilon (z) = A_0 + A_1~(1+z) + A_2~(1+z)^2~.\lb{wp}
\ee
By definition, $\epsilon (0)=1$, hence $A_0 + A_1 + A_2=1$. Note that we have for (\ref{wCPL})
\be
\epsilon (z)= (1+z)^{3(\alpha + \beta)}{\rm e}^{-3\beta \frac{z}{1+z}}~.
\ee
Actually, rather than delimiting some restricted bounded domain in the parameter space,
some observations on small redshifts $z\lesssim 0.35$ single out a prefered direction.
Variations along this direction are essentially unconstrained while variations normal to
it are most efficiently constrained. This is the case for baryon oscillations data which
constrain $h(z)$ and therefore $w(z)$ as follows
\be
{\Omega_{m,0}}^{1/2}{h(z_{1})}^{-1/3}{\left[\frac{1}{z_{1}}\int_{0}^{z_{1}}\frac{dz}{h(z)}\right]}^{2/3}
             \le 0.469 \pm 0.017~, \lb{oscb}
\ee
with $z_{1}=0.35,~\Omega_{m,0}=0.3$.

When we use the parametrization (\ref{wCPL}), eq.(\ref{oscb}) translates into the constraint at
the 1-$\sigma$ level in the parameter plane $\alpha,~\beta$ (equivalently $w_0,~w_1$),
\be
\alpha + 0.112~\beta = 1+w_{DE,0} + 0.112~w_1 = 0.23 \pm 0.20~.\lb{Cosc}
\ee
As said in the Introduction, we see that these constraints allow for a $w_0$ close to (and possibly
slightly lower than) $-1$ if we take a constant equation of state, while a $w_0$ significantly lower
than $-1$ is allowed but it requires a steeply increasing $w$ near $z=0$ in agreement with recent
analysis of the data.
It is natural to use the constraint (\ref{Cosc}) when we consider the small
$z$ expansion of our scalar-tensor model quantities and of our fit (\ref{wCPL}).

\begin{figure}[ht]
\begin{center}
\psfrag{F}[][][1.5][-90]{$F$}
\psfrag{z}[][][1.5][0]{$z$}
\includegraphics[angle=-90,width=1\textwidth]{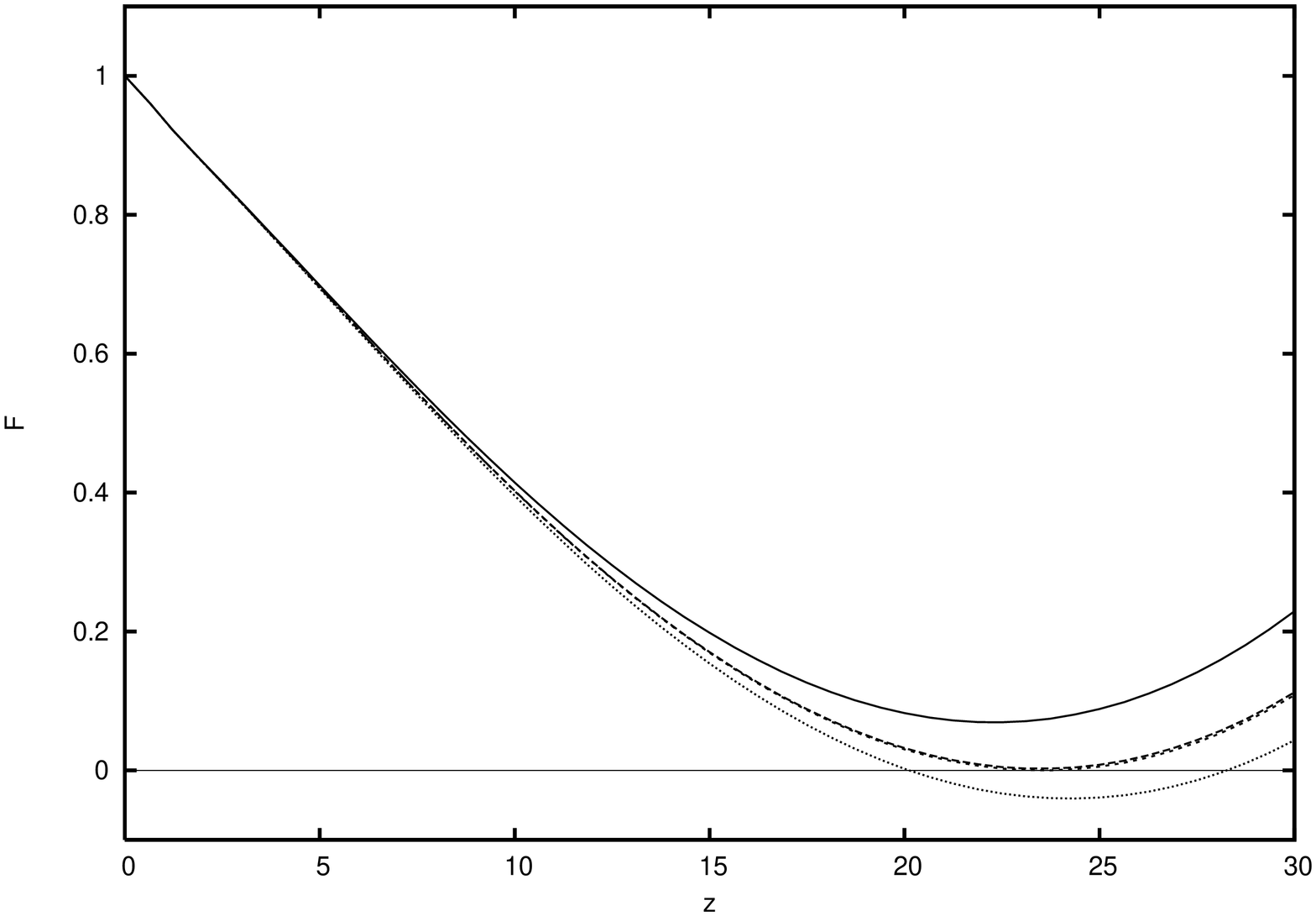}
\end{center}
\caption[]{The quantity $\frac{F(z)}{F_0}$ is displayed for the parametrization (\ref{wCPL}) with
$\alpha\equiv 1+w_0=-0.2$ and $\beta \equiv w_1=0.4$ and $F_1=0$. We have the following values
for $\frac{\Omega_{U,0}}{\Omega_{DE,0}}=\frac{\Omega_{U,0}}{0.7}$ from bottom to top:
$0.9758,~0.975824492,~0.975826,~0.97587$.
The second curve has its minimum at $F=0$ and is superimposed on the third curve which has its
minimum at $F=2.4\times 10^{-3}$.}
\end{figure}

\begin{figure}[ht]
\begin{center}
\psfrag{phi}[][][1.5][-90]{$\varphi'^{2}$}
\psfrag{z}[][][1.5][0]{$z$}
\includegraphics[angle=-90,width=1\textwidth]{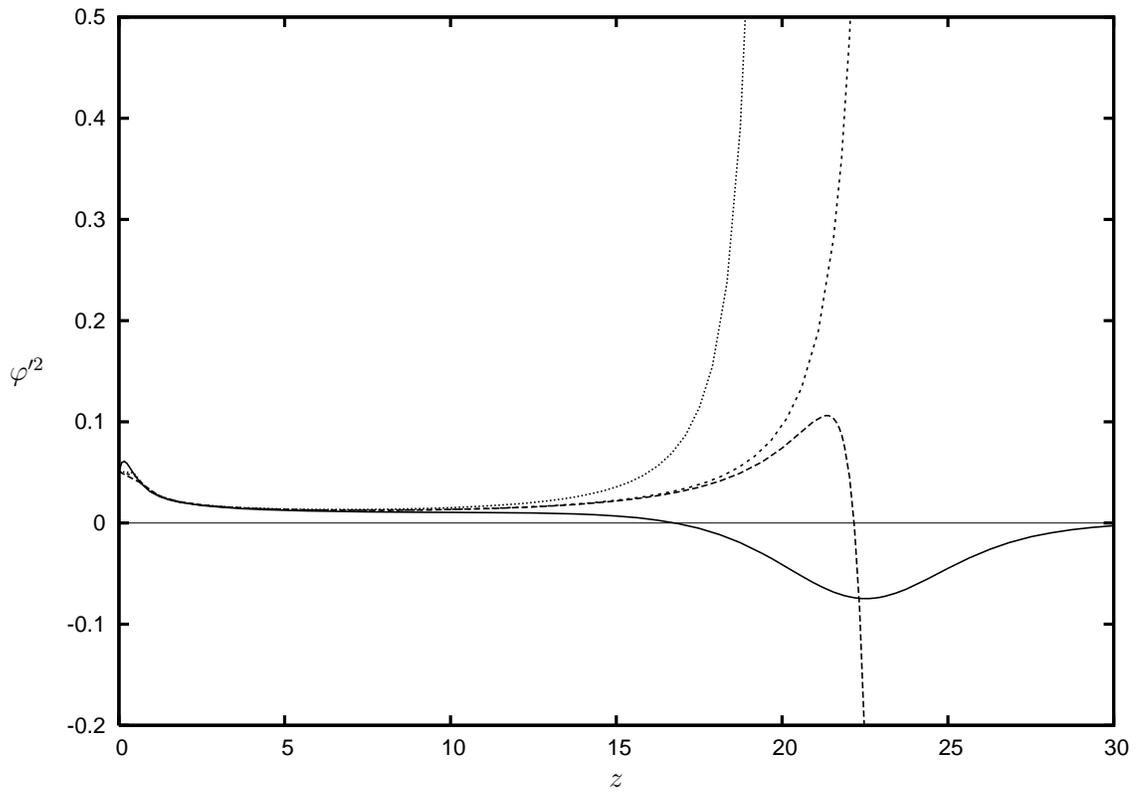}
\end{center}
\caption[]{The quantity $\phi'^2$ is shown for the same models as Figure 2. The short, resp.
long, dashed curve corresponds to $\frac{\Omega_{U,0}}{\Omega_{DE,0}}=\frac{\Omega_{U,0}}{0.7}
=0.975824492,~{\rm resp}~0.975826.$}
\end{figure}
\begin{figure}[ht]
\begin{center}
\psfrag{F}[tl][br][1.5]{$F$} \psfrag{z}[][][1.5][0]{$z$}
\includegraphics[angle=-90,width=1\textwidth]{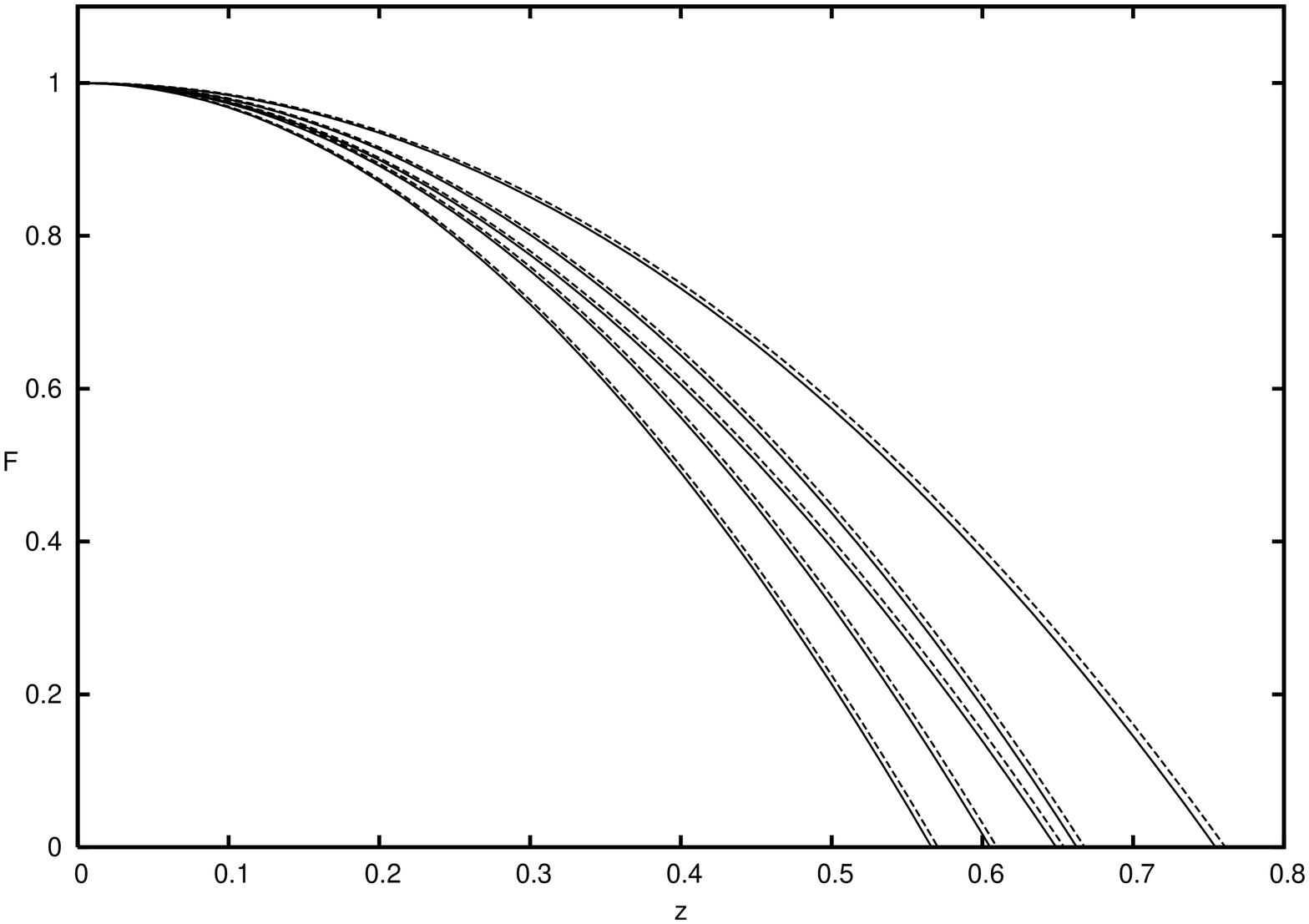}
\end{center}
\caption[]{The function $\frac{F}{F_0}$ is shown for several models with vanishing potential $U=0$.
The solid lines correspond to the initial conditions $F'_0=0$ while the dashed lines
correspond to the maximal values allowed by the solar system constraint
$\omega_{BD,0} > 4 \times 10^4$. We have from left to right the following equation of state
parameter $w$: $-2,-1.5$, polynomial expression (\ref{wpg}), $-1,-0.5$. It is seen that the
limit of regularity of these models corresponds to very low redshifts, $0.566\le z\le 0.663$
for $-2\le w\le -1$. Note that the polynomial expression (\ref{wpg}) represents crossing of
the phantom divide.}
\end{figure}

\section{Reconstruction for a constant potential $U$}
Setting $U=0$, we consider either $w_{DE}<0$ constant, or else $w_{DE}(z)$ of the form
(\ref{wCPL}) or (\ref{wp}), whose variation is parameterized by two parameters. This
extends the investigations in \cite{EP01}, where only the case $w_{DE}=-1$
was considered. The interesting result obtained there was that $F$ vanishes, and hence the
theory is singular, at very low redshifts $z\approx 0.66$ for $F'(0)=0$ and
at only slightly higher redshift $z\approx 0.68$ if one fully exploits the possible initial
conditions allowed by the solar system constraints. It is therefore interesting
to investigate how this result is affected when we take different equations of state.
In particular we will also consider here equations of state with $w_{DE}<-1$.
As SNIa observations point to a varying equation of state which is of the phantom type
on very low redshifts $z\lesssim 0.3$, this case is considered too.
For the polynomial expression (\ref{wp}), for definiteness we will use the best fit to the
``Gold data set'' consisting of 156 supernova which was proposed in \cite{LNP05} (with the
priors $\Omega_{m,0}=0.3$ {\it and} $\Omega_{k,0}=0$):
\be
A_1 = -5.94 \pm 3.61 ~~~~~~A_2 = 2.39 \pm 1.47~.\lb{wpg}
\ee
These values are similar to those earlier obtained in \cite{ASS04} for the same dataset.
As we can see, (\ref{wpg}) has large uncertainties.
The corresponding equation of state is today of the phantom type but rapidly increases to a
positive value $w_{DE}\simeq 0.1$ at $z\simeq 0.8$ and then decreases to its asymptotic value
$w_{DE}=-\frac{1}{3}$ but it is still slightly positive at $z=2$.
We first note from (\ref{cp0a}) that $F_2$ satisfies
\be
F_2 = \frac{3}{2} ~(w_{DE,0} - 1)~\Omega_{DE,0} + 3~\Omega_{U,0}~. \lb{F2}
\ee
Therefore $F_2<0$ for $U=0$ as long as $w_{DE,0}<1$ while $F_2$ decreases with decreasing
$w_{DE,0}$.
The results obtained numerically are shown in Figure 4. It is seen that we keep essentially
the same picture, in
particular $F$ vanishes more rapidly when $w_{DE}$ decreases, hence the problem is even more
severe for phantom DE with constant $w_{DE}$. Even the fit (\ref{wpg}) where $w_{DE}$ varies
substantially at low redshifts yields essentially the same behaviour. The same result as for
(\ref{wpg}) is obtained using the parametrization (\ref{wCPL}) with $\alpha=-0.377$ and
$\beta=2$. This is similar to earlier results showing that a large variation of the equation
of state around $w_{DE}=-1$ on low redshifts can result in essentially the same $d_L(z)$
\cite{CP01,MBMS02} for $z\lesssim 1$ and one can understand from the general expression
eq.(\ref{fsol}) why all cases displayed in Figure 1. will have basically the same behaviour
regarding the evolution of $F(z)$.
The initial condition $F_0'=F_1=0$ and $\Phi'(0)\ne 0$ means physically that the theory corresponds
strictly to General Relativity today, $\omega_{BD,0}^{-1}=0$. The solar system constraints allow
for a very small nonvanishing $F_1$, see eq.(\ref{dF0}), and the corresponding change in $z_{max}$
is very marginal as can be seen on Figure 4. To summarize, looking only at $z=0$ one could think
that models with $w_{DE,0}=-1$ and $U=0$ are allowed, however we see that the cosmological
evolution of such models leads to a singularity at very low redshifts generalizing the results
obtained earlier \cite{EP01}.
It is clear from these results that a scalar-tensor theory of gravity with vanishing potential
$U$ is definitely excluded by the Supernovae data.

The next natural step is to consider constant (nonvanishing) potentials $U$. Of course, one does
not expect such a behaviour to be relevant up to very large redshifts, but it is certainly a
sensible approximation to start with on small redshifts $z\lesssim 1.5$.
We note from (\ref{F2}) that $F_2<0$ for $w_{DE,0}<-1$. By taking $\Omega_{U,0}\approx \Omega_{DE,0}$
one has the smallest possible $F_2$ for given $w_{DE,0}$ and these are the cases which are found to
have the largest range of validity.
We find that many models are allowed which are perfectly viable on small redshifts $z\lesssim 1.5$.
Typically, these models become singular at some higher redshift well beyond $z=2$.
There are several possibilities: either it is the quantity $F$
that is vanishing first or it is the quantity $\phi'^2$ that vanishes first. As discussed in Section 6,
when $F=0$ and $F'\ne 0$, the quantity $\phi'^2$ diverges, $\phi'^2\to \infty$.
Models for which both $F$ and $F'$ vanish together can be considered as a limiting case which gives the
largest possible redshift. If we change very slightly the potential $U$ so that when $F'=0$,
$F>0$, one gets $\phi'^2=0$ earlier. These different possibilities are displayed in
Figures 2,3. We have checked the behaviour of such models with $w_{DE}$ starting  below $-1$
as favoured by SNIa data and with $w_{DE}$ quickly becoming larger than -1.
To summarize, we find that models with $\Omega_{U,0}\to 0.7$ {\it and} an equation of state with
$w_{DE}\to -1$ will become singular at arbitrarily high redshifts.
We should remember that $\Omega_{U,0}=0.7$ together with $w_{DE}=-1$ gives back General Relativity
($F=1$). In all other cases some maximal redshift is found where the model becomes singular.

In all our numerical calculations, we neglected radiation since its
energy density is very small at redshifts of interest. However, even in 
principle its presence cannot prevent the occurence of the singularity
at the moment when $G_{eff}$ changes sign ($\phi'^2=0$ in our case)
whose generic (anisotropic) structure is independent of the matter
equation of state (see [55] in this respect).

\section{Asymptotic stability}

The next step is to consider nonconstant potentials. As we have shown
very generally in Section 5, such models are consistent with DE of the
phantom type today and phantom divide crossing at small redshifts. An
example of the reconstruction of a model with phantom DE today and
phantom divide crossing at $z\approx 0.3$ was presented in \cite{Per05}.
However, in this paper the reconstruction was implemented only for small
redshifts $z\lesssim 2$. As was shown in the previous section, one has
to consider the large-$z$ behaviour of a model, too, to prove its viability.

In contrast to the case with constant or vanishing potential $U(\Phi)$, a growing potential
$U(\Phi)$ allows for the construction of scalar-tensor DE models which are viable for all
redshifts and evolves according to the fit (\ref{wCPL})
\be
w(z) = {\rm const} = w_0 + w_1 = -1 + (\alpha + \beta) > -1 ~~~~~~~~~~~~~~~~~z\gg 1~.
\ee
As a particular example of such a model, let us assume that the scalar-tensor gravity
approaches GR sufficiently fast at the matter-dominated stage and that DE 'tracks'
matter:
\bea
F & \to & F_{\infty} = {\rm const},~~~|\dot F|\ll HF_{\infty},~~~|\ddot F|\ll H^2F_{\infty}, \\
H^2 & \propto &  (1+z)^3~~~~~~~~~~~~~~~~~~~~~~~~~~~~~~~~~~~~~~~~~~~~~~~~~~1\ll z\ll z_{eq}~,\lb{asol}
\eea
where $z_{eq}$ is the redshift at the matter-radiation equality. In this way we recover in
the past the usual behaviour $a\propto t^{\frac{2}{3}}$. However, the constant value
$F_{\infty} < F_0$ may not be too small compared to $F_0$. In order
to satisfy the BBN constraints \cite{CDK04,UIY05}, the following inequality is required
\be
(F_0-F_{\infty})/F_0 < 0.1~, \lb{BBN}
\ee
which can be easily satisfied. Indeed, as can be seen from (\ref{ineq}),
we have for constant $F$ and $U>0,~\phi'^2>0$
\be
F_0 - F_{\infty} < \frac{\rho_{DE}}{\rho_{m} + \rho_{DE}} F_0~.
\ee
On the other hand, the assumption $F_0>F_{\infty}$ provides a good matching to the
small-$z$ expansion derived in Sec.~5 with $F_2<0$ and a very small $F_1$. In view of
(\ref{BBN}), this matching should occur at a sufficiently small $z$, too.

This scaling behaviour of DE corresponds to the asymptotic solution (\ref{asol}) with
$w_0 = -w_1$, equivalently $\alpha + \beta = 1$. As is well known, it can be obtained by
taking an exponential potential $U(\Phi)$ at $\Phi \to \infty$, i.e. at large redshifts:
\be
U \propto {\rm e}^{ \sqrt{ \frac{3}{2 F_{0}\Omega_{U,\infty}} }\Phi} ~,
\ee
%
where $\Omega_{U,\infty}$ is the constant asymtotic value of $\Omega_U$
at $z\gg 1$ which is a free parameter formally. However, actually it 
should be small (less than a few percents) to obtain the correct value of 
the growth factor of density perturbations during the total 
matter-dominated stage. The total DE energy density in terms of the 
critical one $3F_0H^2$ is
\be
\Omega_{DE,\infty} = 2\Omega_{U,\infty} + {F_0 - F_{\infty}\over F_0} <1~.\lb{Omega-infty} 
\ee
Note that the term ${F_0 - F_{\infty}\over F_0}$ in (\ref{Omega-infty})
always has the same equation of state as the main background matter. Thus,
to obtain behaviour different from $w_0=-w_1$ at large $z$ during the matter
dominated stage is possible if $F_0 = F_{\infty}$ only which requires 
additional fine tuning and is not natural. 

\section{Conclusions}

In this work we have considered the viability of scalar-tensor models of Dark Energy. We have
used different types of observations: Solar System constraints which constrain the model
{\it today}, and other data like the supernova data which constrain its cosmological evolution,
in particular the time evolution of the DE equation of state parameter $w_{DE}$. We were
interested specifically in models which violate the weak energy condition on small redshifts
$z\lesssim 0.3$ and in any case today. We have found the formal general integral solution for
$F(z)$ when we reconstruct it for given $H(z)$, which can be obtained from the $d_L(z)$ data,
and for given $U(z)$. This general solution allows immediately for an integral representation
of scaling solutions. We have constructed scaling solutions and shown that they exist in models
with $F=\alpha \Phi^2$ with $\alpha=$constant, in these models the Brans-Dicke parameter
$\omega_{BD}$ is constant. Only for nonzero potentials can these models have scaling solutions
with constant $w_{DE}<-1$. However, it is shown that for these models $|w_{DE}+1|$ is very small
with $|1 + w_{DE}|\sim \omega_{BD}^{-1}$. We have further performed systematically the small
$z$ expansion of the theory and used it to extract various observational constraints. We recover
that a large positive $\omega_{BD,0}$ requires $|F_1|\ll 1$ where $F_1$ is the first derivative
today of $\frac{F}{F_0}$. We find that a significantly phantom DE today ($w_0<-1.01$) implies
that $F_2$, the second derivative today of $\frac{F}{F_0}$, must be negative and not small, thus,
significantly larger than $F_1$ by modulus. However, while necessary this is not a sufficient
condition for phantom DE today, for example for vanishing potential $F_2<0$ whenever $w_{DE,0}<1$.

The condition $|F_2|\sim 1$ while the Solar System data require
$|F_1|<10^{-2}$ (i.e. anomalously small) is the only 'fine tuning' required to get
a significantly phantom DE at the present time in scalar-tensor gravity. 
Note that, since the derivatives $F_i(z)$ are not parameters of the effective 
microscopic Lagrangian (1) but depend on initial conditions in the early Universe too, 
it could be even better to call this a "cosmic coincidence". 
Our point of view is that the condition $|F_2|\sim 1$ cannot be excluded by pure 
thought, only observations will possibly do it: in the absence of this condition 
when all $F_i$ are of the same order as $F_1$, the general prediction is that the 
amount of possible phantomness in scalar-tensor DE models is very small, less than $1\%$.

As for the solar system
constraints, we have shown that the Post-Newtonian parameter $\gamma_{PN,0}$ must satisfy
$\gamma_{PN,0}-1\simeq - \omega_{BD,0}^{-1}< 0$, this is a general requirement for the viability of
our model. On the other hand, a significantly phantom DE today implies $\beta_{PN,0} > 1$. Combining
those results, we find that the negative quantity
$\frac{\gamma_{PN,0}-1}{\beta_{PN,0}-1}$ does not depend on $F_1$ and can be expressed in function
of $F_2$ and $1+w_{DE,0}$. This would enable us to find $F_2$ from the Solar System constraints
provided we know $1+w_{DE,0}$ from other cosmological data. Still for phantom DE today, we find
that ${\dot G}_{{\rm eff},0}$ has the same sign as $F_1$. So, a measurement of the sign
of ${\dot G}_{{\rm eff},0}$ would give us the sign of $F_1$ if we know we have $1+w_{DE,0}<0$.
On the other hand, a measurement of both quantities with opposite signs would rule out phantom DE
today though measuring the sign of $F_1$
%
%
can be hard to determine observationally.
However, due to the already confirmed smallness of $F_1$, connection between cosmological
and Solar System tests of dark energy is rather one-way since it requires much greater accuracy
from the latter ones for the determination whether DE is phantom at present or not. While a positive
detection of $\beta_{PN,0}>1$ is a strong argument for the phantom DE at present, a negative
result (no measurable deviation of $\beta_{PN,0}$ from unity) tells us nothing regarding
DE properties. Also, the Solar System tests are clearly unable to provide $w_1$ in any
reasonable future, so no information about the possibility of the phantom divide crossing may be
expected from them.

We have also considered numerically the reconstruction of $F$ for various $w_{DE}$, including
constant as well as varying equations of state of the phantom type. Generalizing results obtained
in \cite{EP01} for a pure cosmological constant, we find that models with a vanishing potential
(see Figure 3) are ruled out and lead to a singular behaviour for $z<0.66$. While it is clearly
possible to have phantom DE today with $U=0$ without conflicting with the data, the cosmological
evolution of these models rules them out. For models with constant nonvanishing potentials, which
can be considered as a good approximation on small redshifts for more general models with varying
potentials, we find that it is easy to have models in agreement with observations on small
redshifts $z\lesssim 2$. However, it is interesting that these models generically have a maximal
redshift where they become singular. So, to construct a scalar-tensor DE model having a
sufficiently long matter-dominated stage, a non-constant potential $U(\Phi)$ is required, and
we have presented an example of such a model where DE tracks matter at large redshifts.

Therefore, the final conclusion is that the generic scalar-tensor gravity (\ref{L}) with the
two functions $F(\Phi)$ and $U(\Phi)$, derived from some underlying theory (e.g., from brane
cosmology) or taken from observational data, has enough power to provide internally consistent
cosmological models with a temporarily phantom DE (at present, in particular) and with a
regular phantom divide crossing in the course of the evolution of the Universe.

\section{Acknowledgements}
A. A. S. was partially supported by the Russian Foundation for Basic
Research, grant 05-02-17450, and by the Research Program
``Astronomy'' of the Russian Academy of Sciences, and acknowledges 
financial support from Universit\'e Montpellier II where part of this 
work was carried out.

\begin{landscape}
{\bf APPENDIX}
\vskip 20pt
\par\noindent
We summarize in the table below, the integrability of scaling solutions using (\ref{fsol}) for
several exponents $n\equiv 3\gamma$. We use the following abbreviations:
\par\noindent
a.n.i.: analytically non integrable
\par\noindent
Elliptic, log, Hypergeo : solutions expressed in terms of resp., elliptic, log (or $\arg\tanh$) and hypergeometric functions
\par\noindent
$P^{(n)}(x)$: polynomial in $x$ of degree $n$
\
\vskip 20pt
\par\noindent

\begin{tabular}{|c|c|c|c|c|c|c|c|c|c|}
\hline $n $ &  $<0$ & $0$ & $1$ & $2$ & $3$ & $4$ & $5$ & $6$ & $7,\ \ 8,\ \ 9$\\
$w$ &  $<-1$ & $-1$ & $-2/3$ & $-1/3$ & $0$ & $1/3$ & $2/3$ & $1$
& $2/3 , 5/3, 2$\\ \hline \hline $A_{0}$ & a.n.i. & Elliptic &
Elliptic & $\log$ & $-\frac{2}{\sqrt{(1+k)x}}$ &
$\frac{P^{(1)}(x)}{x\sqrt{k+1/x}}$ & Elliptic & Elliptic &
Hypergeo \\
$A_{1}$ & a.n.i. & $\log$ & Elliptic & $\log$ &
$-\frac{2}{3\sqrt{(1+k)x^3}}$ &
$\frac{P^{(2)}(x)}{x^2\sqrt{k+1/x}}$ & Elliptic &
$-\frac{2}{3}\frac{1+kx^3}{\sqrt{x^3+kx^6}}$ & Hypergeo \\
$A_{4}$ & a.n.i. & $\log$ & Elliptic & $\log$ &
$-\frac{2}{9\sqrt{(1+^)x^9}}$ &
$\frac{P^{(5)}(x)}{x^5\sqrt{k+1/x}}$ & Elliptic &
$\frac{2}{9}\frac{(1+kx^3)(-1+2kx^3}{x^3\sqrt{x^3+kx^6}}$ &
Hypergeo \\ \hline \hline $\beta_{1}$ & a.n.i. &  a.n.i. &  a.n.i.
& a.n.i. & $\frac{1}{3(1+k)x^2}$ & $\frac{1}{3x^2}-\frac{4k}{3x}$
& a.n.i. & a.n.i. & a.n.i. \\ $\beta_{2}$ & a.n.i. &  a.n.i. &
a.n.i. & a.n.i. & $\frac{2}{45(1+k)x^5}$ &
$\frac{2}{45x^5}-...+\frac{256k^4}{314x}$ & a.n.i. & a.n.i. &
a.n.i. \\ \hline
\end{tabular}
\ \\ \ \\ \ \\
\newcommand{\ud}{\mathrm{d}}
\[\mathrm{where} \qquad \qquad A_{n} \propto \int\!\!\frac{\ud x}{x^nh(x)} \qquad \qquad
\beta_{1} \propto \int\!\!\frac{\ud x}{h(x)}\int^x\!\!\!\frac{\ud
\eta}{\eta h(\eta)} \qquad \qquad \beta_{2} \propto
\int\!\!\frac{\ud x}{h(x)}\int^x\!\!\!\frac{\ud \eta}{\eta^4
h(\eta)}\qquad \qquad \qquad \phantom{.}\]
\\ \ \\
\end{landscape}
\clearpage

\end{document}